\documentclass[aps,preprint,floats,nofootinbib]{revtex4}
\usepackage{amsmath}
\usepackage{amssymb}
\usepackage{latexsym}
\usepackage{graphicx}

\newlength{\defbaselineskip}
\newcommand{\setlinespacing}[1]%
           {\setlength{\baselineskip}{#1 \newcommand}}

\def\lsim{\mathrel{\raise.3ex\hbox{$<$\kern-.75em\lower1ex\hbox{$\sim$}}}} 
\def\gsim{\mathrel{\raise.3ex\hbox{$>$\kern-.75em\lower1ex\hbox{$\sim$}}}} 
\begin{document}

\preprint{
\hfill
\begin{minipage}[t]{3in}
\begin{flushright}
\vspace{0.0in}
FERMILAB--PUB--09--652--A\\
\end{flushright}
\end{minipage}
}

\hfill$\vcenter{\hbox{}}$

\vskip 0.5cm

\title{Synchrotron Emission From Young And Nearby Pulsars}
\author{Christopher M.~Kelso$^1$ and Dan Hooper$^{2,3}$}
\affiliation{$^1$Department of Physics, University of Chicago \\
$^2$Department of Astronomy and Astrophysics, University of Chicago \\
$^3$Theoretical Astrophysics Group, Fermi National Accelerator Laboratory}
\date{\today}

\bigskip

\begin{abstract}

The rising cosmic ray positron fraction reported by the PAMELA collaboration has lead to a great deal of interest in astrophysical sources of energetic electrons and positrons, including pulsars. In this paper, we calculate the spectrum of synchrotron emission from electrons and positrons injected from 376 young pulsars ($<$\,$10^6$ years) contained in the ATNF catalog, and compare our results to observations. We find that if objects such as the Vela and Crab pulsars have injected $\sim$10$^{48}$ erg or more in energetic electrons and/or positrons, they are expected to produce bright and distinctive features in the synchrotron sky. Intriguingly, we predict hard synchrotron emission from these regions of the sky which is qualitatively similar to that observed by WMAP.

\end{abstract}

\maketitle

\newpage

\section{Introduction}

In polar cap~\cite{ruderman75,arons83} and outer gap~\cite{chenghoruderman86} models, electrons are predicted to be accelerated in pulsar magnetospheres. The resulting electromagnetic cascades can include photons that are capable of producing highly relativistic electron-positron pairs. Through such mechanisms, it is plausible that pulsars generate a significant fraction of the GeV-TeV electrons and positrons present in the cosmic ray spectrum~\cite{pulsar,pulsar2,pulsar3,aharonian,zhangcheng,grimani,middle}. 

Recent observations from the PAMELA experiment and the Fermi Gamma Ray Space Telescope (FGST) have generated a great deal of interest in possible sources of high energy cosmic ray electrons and positrons, including pulsars~\cite{pulsar,pulsar2,pulsar3}. In particular, the PAMELA collaboration has reported that the cosmic ray positron fraction (the ratio of positron to electrons-plus-positrons) increases rapidly between approximately 10 GeV and 100 GeV~\cite{pamela} (see also Refs.~\cite{heat,ams}). This is in striking contrast with the predictions of models in which the cosmic positron flux is dominated by the secondary products of hadronic cosmic ray interactions~\cite{serpico,secondaries}. More recently, the FGST collaboration has published their measurement of the cosmic ray electron (plus positron) spectrum between approximately 20 GeV and 1 TeV~\cite{fgstelectron}. This measured spectrum contains a greater flux of electrons above $\sim$100 GeV than is predicted in typical galactic cosmic ray models~\cite{fgstinter}, although not so much as previously reported by the ATIC collaboration~\cite{atic}. 

Among other plausible interpretations of these observations, it has been suggested that a nearby and relatively young pulsar (or pulsars) may be the source of the excess particles~\cite{pulsar,pulsar2}. Assuming that the responsible astrophysical source is not unique in the Milky Way, other regions of our galaxy are also expected to contain large fluxes of highly relativistic electrons and positrons.  Such particles generate synchrotron photons through their interactions with the Galactic Magnetic Field, as well as inverse Compton emission by scattering with starlight and other radiation. Interestingly, emission consistent with synchrotron and inverse Compton emission from the inner Milky Way has been observed in the data of WMAP~\cite{haze} and FGST~\cite{fermihaze}, respectively, leading to the conclusion that a large flux of highly relativistic electrons/positrons is present in the inner kiloparsecs of our galaxy. 

At this time, the origin (or origins) of the GeV-TeV electrons/positrons observed in the local and inner Milky Way has not been unambiguously determined. Among other suggestions, a great deal of attention has been given to the possibility that the excesses observed by PAMELA, FGST, and WMAP may be the products of dark matter particles annihilating~\cite{ann,fgstdark,hazedark} or decaying~\cite{decay} in the halo of the Milky Way. In particular, if the dark matter consists of particles with a mass between several hundred GeV and several TeV, and annihilates or decays primarily to leptons, their products may be able to account for each of the aforementioned observations~\cite{ann,fgstdark,hazedark}.

In this paper, we consider the synchrotron emission from the electrons and positrons injected from nearby and young pulsars. If such emission could be observed and identified, it would provide a valuable test of the origin of the PAMELA and FGST excesses. In particular, the observation of synchrotron emission from regions around young pulsars (or the lack of such emission) would significantly weaken (or strengthen) the case for annihilating or decaying dark matter being the source of the PAMELA and/or FGST signals.

\section{Synchrotron From Regions Around Young Pulsars}

Energetic electrons and positrons generated by pulsars move through the Galactic Magnetic Field, steadily losing energy through synchrotron and inverse Compton scattering. These processes can be modeled by the propagation equation~\cite{prop}:
\begin{equation}
\frac{\partial}{\partial t}\frac{dn_e}{dE_e} = \frac{\partial}{\partial E_e}\left(bE_e^2\frac{dn_e}{dE_e}\right)-DE_e^{a} \, \nabla^2\frac{dn_e}{dE_e}+Q(E_e,\vec{x}),
\label{eq:transport}
\end{equation}
where $dn_e/dE_e$ is the number density of electrons and positrons per unit energy, $bE_e^2$ is the energy loss rate, $DE_e^a$ is the energy dependent diffusion coefficient, and $Q(E_e,\vec{x})$ is the source term which describes the spectrum and time dependence of injected electrons/positrons.  Throughout this study, we use the following values for the propagation parameters: $a=0.43$ and $DE_e^a=2.92\cdot 10^{28}{\rm cm}^2 \, {\rm s}^{-1} (E_e/{\rm GeV})^a$, which are consistent with the primary-to-secondary ratios observed in the cosmic ray spectrum~\cite{simet}. In estimating the energy loss rate, we adopt a magnetic field density of $\rho_B \approx 0.6$ eV/cm$^3$, and a density of starlight and dust emission given by $\rho_{\rm starlight+dust} \approx 0.7$ eV/cm$^3$. Combined with the cosmic microwave background, these densities lead to $b=1.56\cdot10^{-16}$\,GeV$^{-1}$\,s$^{-1}$.

For the source term, $Q(E_e,\vec{x})$, we assume the injection spectrum to take the form of a power-law with a spectral index of $\alpha=1.5$, and exponentially cut-off above 600\,GeV: 
\begin{equation}
f(z)=z^{-\alpha}e^{-z},
\label{eq:injectionSpectrum}
\end{equation} 
where $z=E/E_c$ and $E_c=600\,$GeV. This spectral shape is motivated in part by a desire to accommodate the rising positron fraction observed by PAMELA~\cite{pulsar}, and by the pulsar emission model of Ref.~\cite{zhangcheng}. The source term for an individual pulsar is then given by
\begin{equation}
Q(E_e,\vec{x})=\frac{E_{\rm tot}}{E_c^2\sqrt{\pi}}f(z)g(t)\delta^3(\vec{x}),
\label{eq:sourceTerm}
\end{equation} 
where $E_{\rm tot}$ is the total energy injected by the pulsar in electrons/positrons, and $g(t)$ describes the time dependence of the injection (normalized such that $\int g(t) dt=1$).

The energy budget for a pulsar is simply given by its initial angular momentum,
\begin{equation}
E_{\rm tot} < \frac{1}{2}I \Omega^2_0 \approx 10^{49} \, {\rm erg} \, \bigg(\frac{M_s}{1.4 M_{\odot}}\bigg) \bigg(\frac{R_s}{15 \, {\rm km}}\bigg)^2 \, \bigg(\frac{70 \,{\rm ms}}{P_0}\bigg)^2,
\end{equation}
where $M_s$ and $R_s$ are the mass and radius of the neutron star, and $\Omega_0$ is its initial period. It is generally expected, however, that only a relatively small fraction of a pulsar's total spindown power will be transferred into energetic, escaping electrons/positrons. If $\sim$$10\%$ of the spindown energy budget is expended in this way by a young and nearby pulsar (such as B0656+14 or Geminga), it could potentially accommodate the rising positron fraction observed by PAMELA~\cite{pulsar}. Observations of the Vela pulsar also suggest that this is a plausible estimate~\cite{48}. With these motivations in mind, we adopt $E_{\rm tot} = 10^{48}$ erg as our canonical estimate throughout this study.

The function $g(t)$ accounts for the time dependence of the injected electron/positron spectrum. We model this using two different functions.  In the first model, we assume that all of the electrons and positrons are injected at the birth of the pulsar.
%
%, such that $g(t)=\delta(t+\tau)$, where .  If the current time is $t=0$ and the pulsar has a current age of $\tau$, then $g(t)=\delta(t+\tau)$, where $\tau_0$ is the spindown time, given by
%
Alternatively, we consider the case in which the electrons/positrons are injected over an extended period of time during which the pulsar transfers its angular momentum, or spindown energy, into cosmic rays. In this model, the luminosity in electrons/positrons scales with $g(t) \propto (1+t/\tau_0)^{-2}$, where $\tau_0$ is the characteristic spindown time, given by~\cite{pulsar}
\begin{equation}%
\tau_0=\frac{3c^3I}{B_s^2R_s^6\Omega_0^2} \sim 10^4 \, {\rm years}\, \bigg(\frac{2 \times 10^{12}\, {\rm G}}{B}\bigg)^2 \bigg(\frac{M_s}{1.4 M_{\odot}}\bigg) \bigg(\frac{15 \, {\rm km}}{R_s}\bigg)^{4} \, \bigg(\frac{P_0}{70 \,{\rm ms}}\bigg)^2,
\end{equation}
where $B$ is the magnetic field at the surface of the star. In our calculations, we adopt a representative value of $\tau_0=10^4\,$years. In the remainder of this paper, we will refer to the two models of $g(t)$ as the instantaneous injection and spindown models.

%comes from an analysis in \cite{HooperPulsar} where energetics arguments are used to construct the total energy emitted by a pulsar as a function of time after its birth. This paper treats the spindown power of the pulsar as coming solely from magnetic dipole radiation.  Applying this formulation to the current model gives 
%
%\begin{equation}
%g(t)=\theta(t+\tau)\frac{\tau_0}{\left(t+\tau+\tau_0\right)^2},
%\label{eq:gHooper}
%\end{equation}  
%with $\tau_0=3c^3I/\left(B_s^2R_s^6\Omega_0^2\right)$.  In the expression for $\tau_0$, $I=(2/5)M_sR_s^2$ is the moment of inertia of the pulsar of with mass $M_s$, radius $R_s$, initial spin frequency $\Omega_0$, and surface magnetic field $B_s$.  As in \cite{HooperPulsar}, we used a representative value of $\tau_0=10^4\,$years in our models. 
%

To calculate the spectrum of electrons and positrons as a function of time and the distance from a pulsar, we have used the Green's function as given in Ref.~\cite{Pohl}. For the case of instantaneous injection, we were able to obtain an analytic expression for the electron/positron spectrum:
\begin{equation}
\frac{dn_e}{dE_e}=\frac{E_{\rm tot}}{\pi^2E_c^2\,r_0^3}\,x_m^{-2}\,f(z\,x_m)\,\left[h(z,x_m)\right]^{3/2}\exp\left[-\lambda^2h(z,x_m)\right],
\label{eq:dndEdelta}
\end{equation}
where $h(z,x)=z^{1-a}/(1-x^{a-1})$, and $x_m=1/\left(1-z\xi\right)$. We have adopted the scaling parameters $\xi=bE_c\tau$, and $\lambda=r/r_0$, where $r_0^2=4DE_c^a/\left[(1-a)bE_c\right]$, $\tau$ is the age of the pulsar, and $r$ is the distance to the pulsar. 
%Note that our length scale, $\lambda$, differs from that used in Ref.~\cite{Pohl} by $\lambda=\sqrt{\rho}$.  

For the spindown model the integral that gives the electron spectrum is given by
\begin{equation}
\frac{dn_e}{dE_e}=\frac{E_{\rm tot}}{\pi^2E_c^2\,r_0^3}\,\frac{z^{-1}}{bE_c}\int_1^{x_m}dx\, f(z\,x)\,\left[h(z,x)\right]^{3/2}\exp\left[-\lambda^2h(z,x)\right]\,g\left(\frac{1-x}{xz\xi}\tau\right).
\label{eq:dndEHooper}
\end{equation}

To calculate the spectrum of synchrotron emitted by these electrons/positrons, we need to calculate the electron column density over a line-of-sight, $d\sigma_e/dE_e$.  From Eqns.~\ref{eq:dndEdelta} and \ref{eq:dndEHooper}, we see that in both models the distance dependence is found within the same $\lambda^2$ term inside of the exponential. These integrals can be evaluated yielding error functions. The column density is thus found by making the following replacement in Eqns.~\ref{eq:dndEdelta} and \ref{eq:dndEHooper}:
\begin{equation}
\exp\left(-\lambda^2h\right)\longrightarrow\frac{r_0\sqrt{\pi}}{2\sqrt{h}}\exp\left(-\lambda_d^2\,h\,\sin^2\delta\right)\left[1+{\rm erf}\left(\lambda_d\sqrt{h}\cos\delta\right)\right],
\label{eq:columnDensity}
\end{equation}
where $\delta$ is the angle observed, and the distance to the pulsar, $d$, is related to the scaled distance by $\lambda_d=d/r_0$.

%\subsection{Power Spectrum}

At a frequency, $\nu$, the synchrotron power emitted by the electrons along a line-of-sight (per area, per solid angle, per frequency) is given by 
\begin{equation}
\frac{dP}{d\nu}=\frac{27\sqrt{3}\,b\, m_e^4\nu}{128\pi^2\,\nu_B^2\,E_c}\int_0^{1/\xi}dz \, z^{-2}\frac{d\sigma_e}{dE_e}\,u\left(\frac{m_e^2\,\nu}{z^2E_c^2\,\nu_B}\right),
\label{eq:powerSpectrum}
\end{equation}
where $\nu_B=3eB/\left(4\pi m_ec\right)\approx 21\left(B/5\,\mu{\rm G}\right)\,$Hz (we have used $B=5\mu$G throughout this study), and the function $u(x)$ is given by~\cite{dPdv}
\begin{equation}
u(x)=\int_0^{\pi}d\alpha\int_{x/\sin\alpha}^\infty dy\,K_{5/3}(y).
\label{eq:uofx}
\end{equation}

In the following section, we will apply this formalism to calculate the spectrum of synchrotron emission from the known nearby and young pulsars contained in the ATNF database.
       
\section{Results}

In this section, we consider known pulsars and estimate the synchrotron emission from the electrons and positrons they generate. In particular, we consider the pulsars contained in the ATNF pulsar database~\cite{PulsarDatabase}, focusing on those which are relatively young, and thus most likely to generate the brightest signals on the sky. In particular, we include in our calculations the 376 pulsars in the ATNF catalog that are less than one million years old. Electrons and positrons from older pulsars will have lost a significant fraction of their energy, and will have been diluted over large volumes of space, making them unlikely to provide a very bright or distinctive signal.

\begin{figure}[!t]
	\centering
\includegraphics[width=0.92\textwidth]{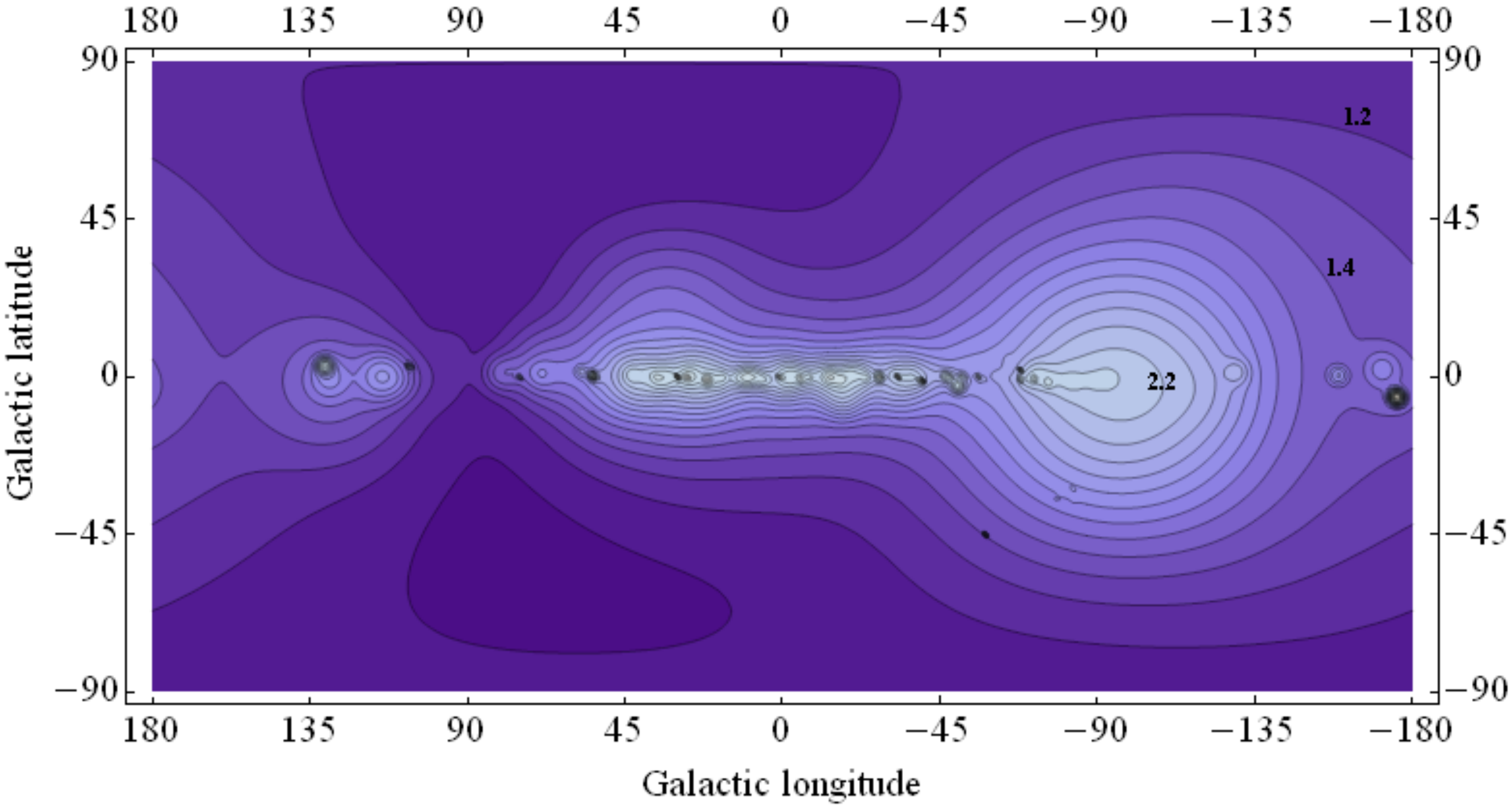}\\
\vspace{0.3cm}
\includegraphics[width=0.92\textwidth]{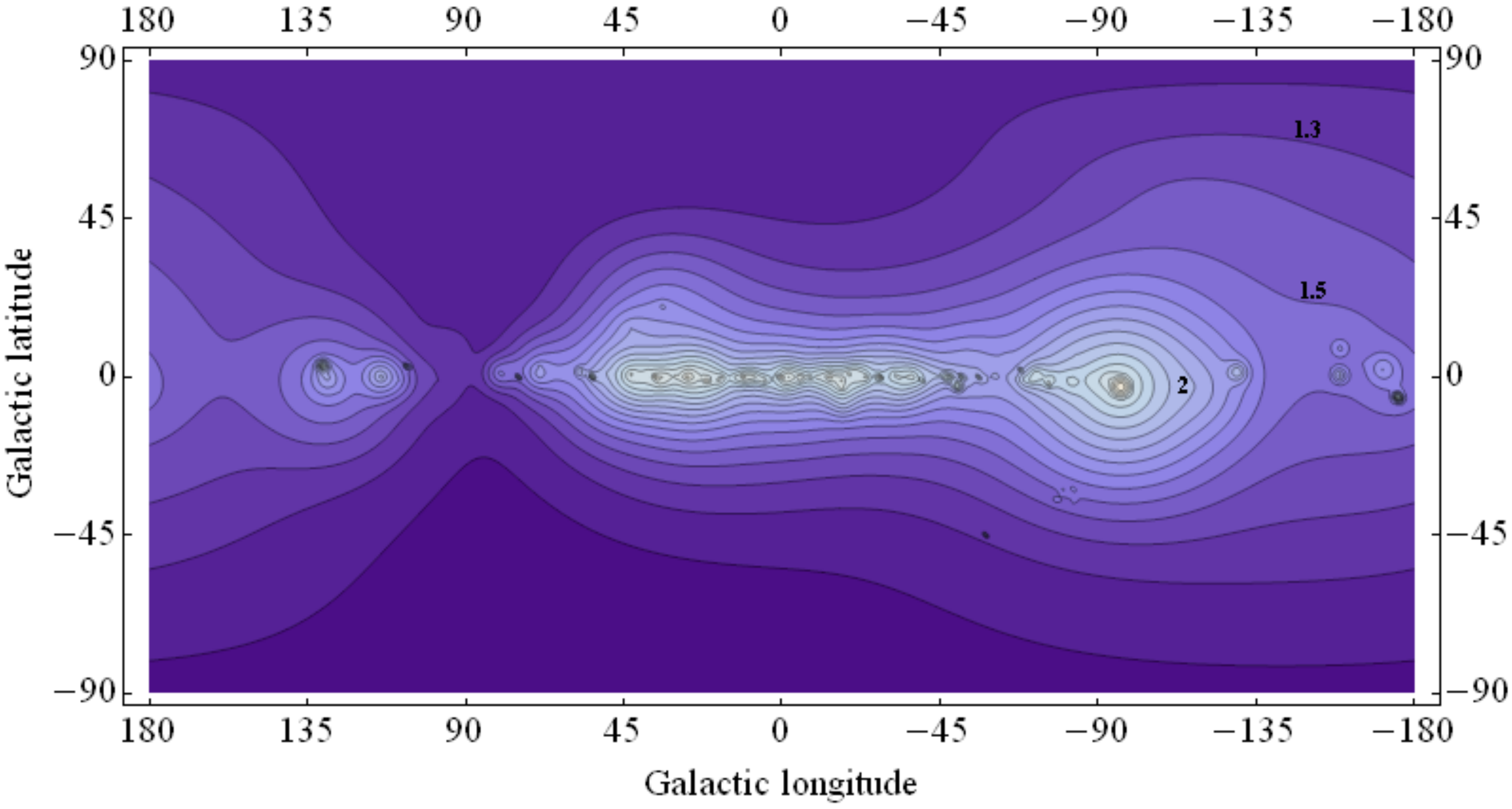}
	\caption{Maps in galactic coordinates of the synchrotron intensity in WMAP's 23 GHz band from the 376 pulsars in the ATNF database less than $10^6$ years old. The upper and lower frames show results assuming instantaneous injection, and a more gradual spindown injection (over a timescale of $\tau_0=10^4$ years), respectively. The plot uses a log-scale in Jy/sr with 0.1 between each contour, and with representative contours labeled (2.0 denoting $10^{2.0}$ Jy/sr, for example). The normalization assumes that each pulsar generates $10^{48}$ erg in electrons and/or positrons. See text for more details.}
	\label{fig:contour22}
\end{figure}

In Fig.~\ref{fig:contour22} we show contour maps of the predicted synchrotron emission from the known catalog of pulsars, calculated as described in the previous section. To aid in the comparison with the WMAP data, we have convolved the synchrotron spectrum with the sensitivities of the relevant WMAP detectors. For a given frequency band of WMAP, we calculate the specific intensity:
\begin{equation}
I_\nu=\int{d\nu\frac{dP}{d\nu}w(\nu)}
\label{eq:convolve}
\end{equation}  
where $w(\nu)$ is the sensitivity of the detectors normalized to unity. The maps in Fig.~\ref{fig:contour22} show the signal in WMAP's $23\,$GHz band for the cases of instantaneous injection (top) and spindown injection over a timescale of $\tau_0 = 10^4$ years (bottom). Overall, the two maps are quite similar, although they vary significantly in the regions near very young pulsars (such as the Vela or Crab pulsars).

\begin{figure}[!t]
	\centering
\includegraphics[width=0.93\textwidth]{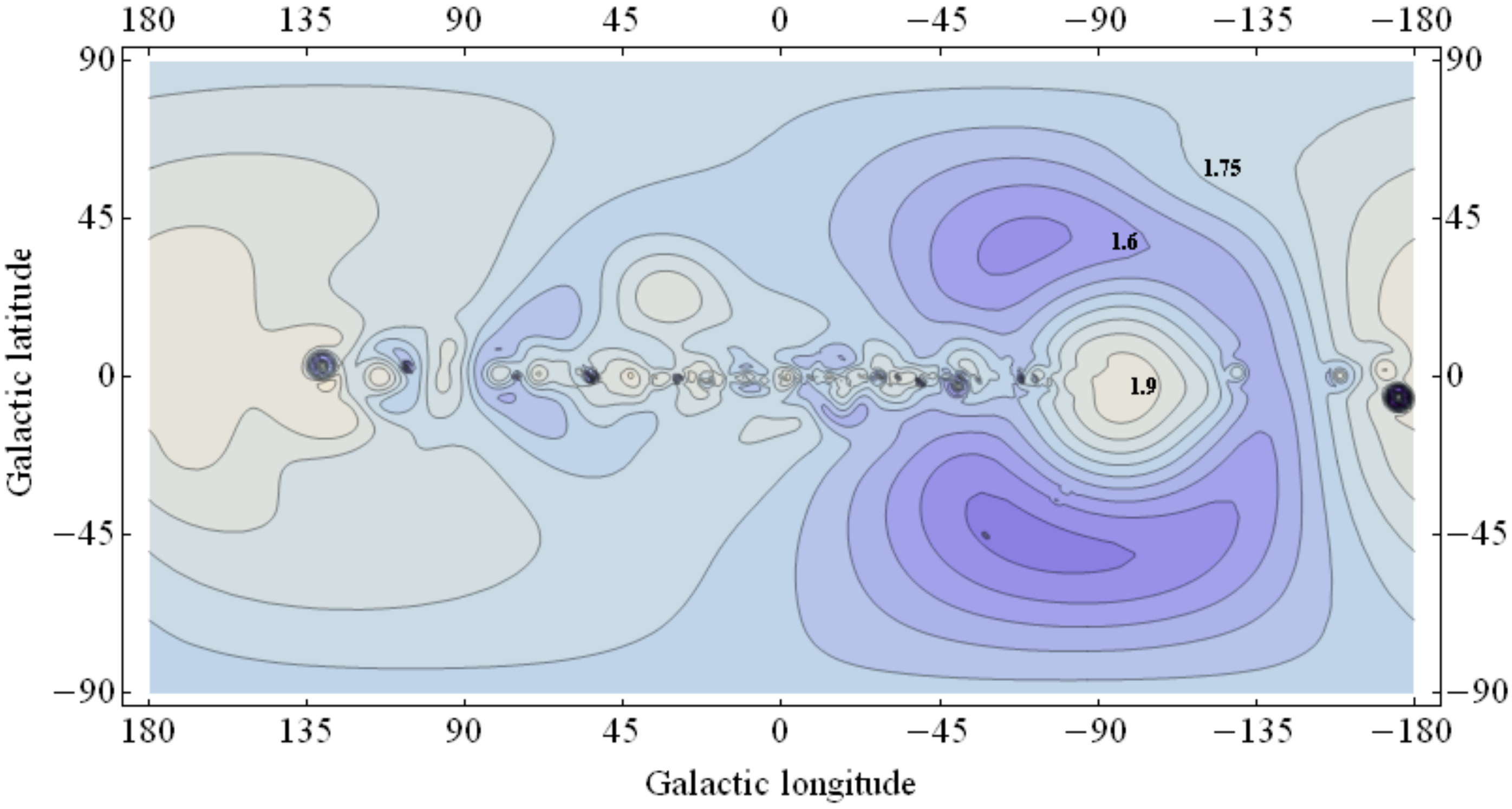} \\
\vspace{0.3cm}
\includegraphics[width=0.93\textwidth]{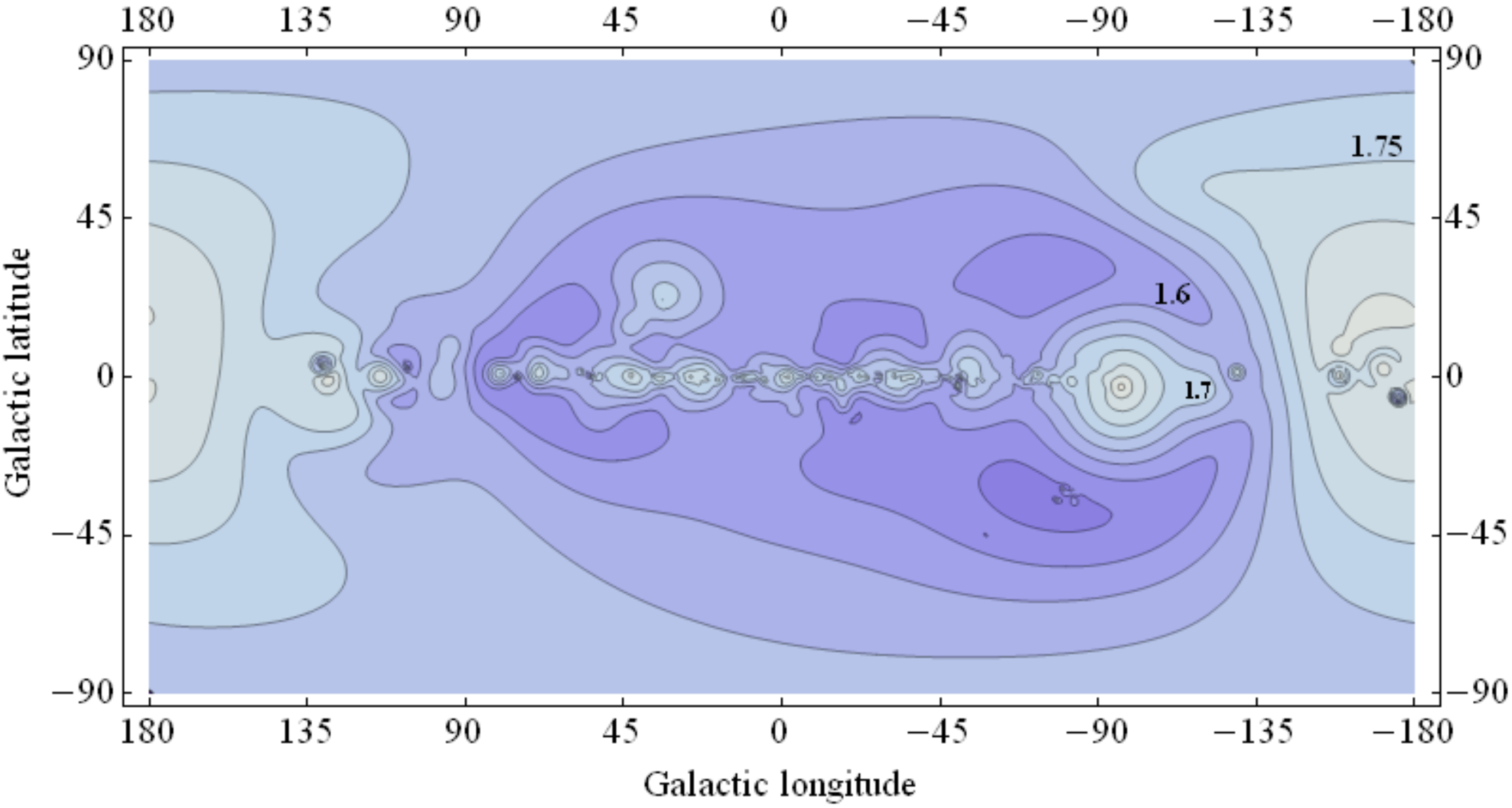}
	\caption{Maps in galactic coordinates of the synchrotron spectral index from the 376 pulsars in the ATNF database which are than $10^6$ years old. The upper and lower frames show results assuming instantaneous injection, and a more gradual spindown injection, respectively. In the upper frame, the contours run from 1.25 (dark regions) to 1.95 (light regions) in steps of 0.05. In the lower frame, they run from 1.5 to 1.9.}
\label{fig:index}%
\end{figure}

In Fig.~\ref{fig:index}, we plot the spectral index of the synchrotron emission from pulsars in our two models.  To obtain the spectral indices shown, we calculated the 23 and 33 GHz maps in the manner described above and, assuming a power-law like form ($I_{\nu}\propto \nu^{-\alpha}$), we then calculate the value of $\alpha$ from the ratio the two specific intensities. Note that the spectral index, $\gamma$ (such that $dN/dE \propto E^{-\gamma}$), is related to $\alpha$ according to $\gamma=2\alpha+1$.

At this point, a few comments are in order. Firstly, some of the most distinctive features seen in our contour maps can be associated with specific young and nearby pulsars. In particular, the emission from the region of the Gum Nebula (which contains the Vela pulsar, at $l=-96.4^{\circ}$, $b=-2.8^{\circ}$) dominates a significant fraction of the sky. A small region around the Crab pulsar ($l=-175.4^{\circ}$, $b=-5.8^{\circ}$) is also clearly present. 

To understand why emission from the Vela and Crab pulsars appears so prominently in the synchrotron sky, consider the angular extent of the synchrotron emission from an individual pulsar, which is roughly given by
\begin{equation}
\theta \sim \frac{D_{\rm dif}}{d} \sim 30^{\circ} \times \bigg(\frac{300 \, {\rm pc}}{d}\bigg) \bigg(\frac{\tau}{10,000 \,\, {\rm years}}\bigg)^{1/2},
\end{equation}
where $D_{\rm dif}$ is the diffusion distance scale, $\tau$ is the age of the pulsar, and $d$ is the distance to the pulsar. Of those pulsars in the ATNF catalog within 3 kpc (approximately 10 times the distance to Vela), there are only 6 pulsars with an angular extent of synchrotron emission smaller than produced by Vela. Of these, all but the Crab pulsar are considerably more distant, and thus less bright in synchrotron.  Other nearby pulsars are bright, but too extended to provide a distinctive synchrotron signal. The Geminga pulsar, for example, is roughly as close to the Solar System as Vela, but is roughly 30 times older. As a result, its electrons fill a volume which encompasses the Solar System, making it difficult to detect its synchrotron emission on the sky, despite being one of the brightest sources of gamma-rays~\cite{FGSTcatalog}.  Furthermore, very young pulsars other than Vela are in some cases too compact to distinguish from other emission from sources near the Galactic Plane.

From Fig.~\ref{fig:index}, one can also notice that the spectral index of synchrotron emission from young, bright pulsars varies with location. Around the Vela Pulsar ($l=-96.4^{\circ}$, $b=-2.8^{\circ}$), for example, the spectrum is harder farther away from the source, which is a result of the fact that the most energetic electrons propagate away from the pulsar more quickly than its lower energy counterparts, leading to a particularly hard electron spectrum in the region of the surrounding ring.

Ideally, one would directly compare results such as those shown in Figs.~\ref{fig:contour22} and~\ref{fig:index} with observations of the $\sim$10-100 GHz sky. For example, we can compare our results to the observations of the WMAP experiment. Comparing our maps to the WMAP residuals (the emission not correlated with known foregrounds) shown in Fig.~7 of Ref.~\cite{doblerfinkbeiner}, for example, we notice a number of features in common with our results. In particular, the most distinctive region of our maps (the region around Vela) is also one of the brightest in the WMAP residual maps (with residual emission on the order of $\sim$1 kJy/sr, which is not radically different from the $\sim$0.2 kJy/sr found in this region of our map for a normalization of $10^{48}$ erg having been emitted from the Vela pulsar in electrons/positrons). Furthermore, this region possesses a notably hard spectral index, consistent with that predicted from cosmic electrons produced by the Vela pulsar over the past 11,000 years. In Fig.~7 of Ref.~\cite{doblerfinkbeiner}, one can also notice a small and bright region in the vicinity of the Crab pulsar ($l=-175.4^{\circ}$, $b=-5.8^{\circ}$). As these represent particularly complex astrophysical regions, however, one should be careful to not draw overly strong conclusions regarding the nature of these residual emissions.

The spindown time and the quantity of energy emitted into electrons and positrons is expected to vary from pulsar-to-pulsar, potentially impacting the resulting intensity and spectral index maps. For those pulsars which we have found to provide the most distinctive signals (the Vela and Crab pulsars), however, our canonical values are appropriate estimates. Assuming a 1.4 solar mass neutron star of radius 10-15 km, the 33 msec period of the Crab pulsar provides $(2-5)\times 10^{49}$ erg of energy (and somewhat more if its initial period were shorter), with an estimated spindown time of several thousand years. In the case of the Vela pulsar, its observed period of 89 msec again leads to an estimated total energy on the order of $10^{49}$ erg, and a spindown time on the order of $10^4$ years.

The region in which our maps most significantly depart from those shown in Ref.~\cite{doblerfinkbeiner} is in that surrounding the Galactic Center. The residual emission in this region, known as the ``WMAP Haze'' is likely hard synchrotron from a population of highly relativistic electrons/positrons. This is the topic of the following section.

\section{Comments on Pulsars as the Source of the WMAP Haze}

The residual emission known as the ``WMAP Haze'' has a number of intriguing characteristics.  Firstly, it is consistent with being synchrotron emission from a population of electrons/positrons with very hard spectral index, $dN_e/dE_e \propto E^{-\gamma}$, $\gamma \sim 1.8$. This is much harder than expected from supernova remnants (after the effects of propagation and energy losses are taken into account), for example. Efforts to identify the WMAP Haze with other possible emission mechanisms (thermal dust, spinning dust, or soft synchrotron as traced by low frequency maps) have proven unsuccessful. Furthermore, as opposed to correlating to the disk or ISM of the Milky Way, the excess haze emission is approximately radially symmetric with respect to the Galactic Center, with an extent of $\sim$20$^{\circ}$. These characteristics have motivated the possibility that the haze emission is synchrotron from high energy electrons/positrons produced in dark matter annihilations taking place in the inner kiloparsecs of the Milky Way~\cite{hazedark}.

Alternatively, it has been previously suggested that a large population of (several thousand) pulsars in the inner galaxy may collectively produce the electrons and positrons leading to the observed synchrotron haze~\cite{Zurek}. In that study, the authors argue that the very hard observed spectrum can be accommodated with an injected spectral index similar to that considered in this paper, assisted in part by the diffusion hardening that occurs as a result of the propagation model. Furthermore, for appropriate choices of the model describing the spatial distribution of magnetic fields, they show that the morphology of the pulsars' synchrotron emission could be sufficiently similar to radially symmetric to avoid any obvious inconsistency with the observed haze~\cite{Zurek}). Note that our maps do not contain a particularly bright haze-like ({\it ie.}~radially symmetric) signal, in part because our sample of known pulsars does not contain a large fraction of those located in the inner Milky Way. A more complete pulsar catalog would presumably yield a more significant synchrotron flux from the inner Galaxy.

An alternative possibility for the origin of the WMAP haze is a single nearby and young pulsar, aligned along the line-of-sight toward the Galactic Center. A Vela-like pulsar ($\tau \sim 10^4$ years, $d\sim 10^2$ pc), for example, could yield the observed (radially symmetric) morphology and hard spectral index, if located in the desired direction of the sky. To normalize to the observed synchrotron flux, however, would require the pulsar to inject $\sim 10^{49}$ erg into energetic electrons/positrons, making it difficult to understand how such an object would have gone undetected in other wavelengths. We thus conclude that this scenario is very unlikely to explain the presence of the WMAP Haze.

%We also used our model to investigate the possibility that a nearby, unresolved pulsar between the earth and the galactic center could be the source of the WMAP haze.  For this model, we calculated the signal from a single pulsar at a given distance and age but kept the total energy in electrons for the pulsar as an unknown parameter.   The value for the total energy in electrons was determined by finding a $\chi^2$ best-fit for the pulsar signal to the WMAP haze data.
%
%Figure \ref{fig:singlePulsar} shows a contour map for the total energy required in electrons to produce the WMAP haze as a function of the distance and age.  From this plot we can see that the pulsars must be very close (less than about 200\,pc) and young (less than ???10,000 years) to ensure that the total energy is less than our threshold of $10^{49}\,$erg.  With this fairly small window in parameter space, this analysis seems to rule out a nearby, unresolved pulsar as a likely source for the WMAP haze.

\section{Conclusions}

In this paper, we have calculated the intensity and spectrum of synchrotron emission from the highly relativistic electrons and positrons produced in relatively young pulsars. In particular, we considered the 376 pulsars in the ATNF catalog which are less than $10^6$ years old, and produced maps predicting the resulting synchrotron emission. 

If pulsars inject $\sim$$10^{48}$ erg or more into $\sim$10-1000 GeV electrons/positrons, as would be required for such an object to generate the rising positron fraction observed by the PAMELA experiment, they could also produce bright and potentially observable features in the synchrotron sky. This is especially true in the case of very nearby and young pulsars (such as the Vela or Crab pulsars). In the region of the sky within $\sim$30$^{\circ}$ of the Vela pulsar, for example, we predict considerably brighter and harder synchrotron emission than in other regions of the sky away from the Galactic Plane. A smaller region of very bright and hard synchrotron emission is also predicted from the region around the Crab pulsar. 

Intriguingly, the features predicted by our calculations are qualitatively similar to those observed in the WMAP residual maps (the emission not correlated with other known foregrounds)~\cite{doblerfinkbeiner}.  Given the astrophysically complex nature of these regions (the Gum Nebula and Crab Nebula), however, it is not possible at this time to positively identify the origin of this emission.

\bigskip
\bigskip
This work has been supported by the US Department of Energy, including grant DE-FG02-95ER40896, and by NASA grant NAG5-10842.

\end{document}